# Distributed Cooperative Formation Control of Nonlinear Multi-Agent System (UGV) Using Neural Network

Si Kheang Moeurn

School of Automation, Beijing Institute of Technology, Beijing, 100081, China
E-mail: sikheang2498@163.com

**Abstract:** The paper presented in this article deals with the issue of distributed cooperative formation of multi-agent systems (MASs). It proposes the use of appropriate neural network control methods to address formation requirements (uncertainties dynamic model). It considers an adaptive leader-follower distributed cooperative formation control based on neural networks (NNs) developed for a class of second-order nonlinear multi-agent systems and neural networks Neural networks are used to compute system data that inputs layer (position, velocity), hidden layers, and output layer. Through collaboration between leader-follower approaches and neural networks with complex systems or complex conditions receive an effective cooperative formation control method. The sufficient conditions for the system stability were derived using Lyapunov stability theory, graph theory, and state space methods. By simulation, the results of this study can be obtained from the main data of the multi-agent system in formation control and verified that the system can process consistency, stability, reliability, and accuracy in cooperative formation.

## 1 Introduction

Cooperative formation work in the realm of multi-agent systems has demonstrated a plethora of advantages, encompassing notable traits such as superior flexibility performance, extensive sensing coverage, and robustness. This approach has yielded considerable achievements in diverse domains, including but not limited to unmanned ground vehicles (UGVs), unmanned aerial vehicles (UAVs), mobile robots, and autonomous underwater vehicles[1][2]. The central challenge in this context revolves around the effective formulation and maintenance of desired formations through proficient control methodologies. Traditionally, three predominant control frameworks specifically, leader-follower-based, virtual structure-based, and behavior-based strategies have been extensively explored and employed in the field of formation control[3]. Nevertheless, each of these frameworks is not without its inherent limitations. In response to these challenges, researchers have sought alternative solutions, leading to the integration of consensus control strategies[4]. Consensus control strategies, renowned for their robustness and high flexibility, have garnered significant attention. Recognizing their efficacy, a growing number of researchers have extended the application of consensus theory to the domain of formation control, yielding numerous noteworthy results[5][6].

Nonetheless, it is imperative to underscore that the formations observed in the majority of existing results are predominantly time-invariant, a characteristic that evidently falls short of meeting diverse practical requirements. These encompass challenges such as avoiding moving obstacles, adapting to changing environments, and executing other intricate tasks[7]. In response to these limitations, various studies have been conducted to address the realm of time-varying formation control. One notable investigation delved into time-varying formation problems within the context of linear multi-agent systems featuring switching directed topologies. The study established necessary and sufficient conditions for asymptotic stability, offering valuable insights into the challenges posed by dynamic formations. Moreover, in scenarios where formations need to dynamically track the trajectory generated by a virtual or real leader to accomplish specific tasks, the time-varying formation tracking problem emerges. This entails a group of agents maintaining the expected time-varying formation while concurrently tracking the leader's trajectory[8].

It is important to note that the majority of control strategies for formation control rely on accurate states of the leader, encompassing both position and velocity information. However, in certain scenarios, such as employing ultra-wideband localization techniques, obtaining precise position information may be straightforward, while accurate velocity measurements prove challenging due to potential contamination by environmental noise[9][10]. These factors can adversely impact the performance of formation control. To address these challenges, several studies have investigated formation control based on velocity observers.

Moreover, in practical scenarios with distributed control properties, only a subset of agents may have access to the complete state information of the leader. Consequently, it becomes necessary for each agent to reconstruct the leader's states through observer approaches to facilitate simplified control design. Results based on distributed observers have been obtained[11][12]. It is noteworthy that the existing observers are predominantly asymptotically stable or finite-time stable. Recognizing the significance of settling time in formation performance, finite-time convergence, known for its effective pursuit of convergence rates, is considered. As an extension of finite-time stability, fixed-time stability ensures that the settling time function derived from stability analysis is independent of initial conditions and uniformly bounded[13][14].

In addressing uncertainties stemming from parametric uncertainties, modeling errors, and external disturbances,

formation tracking control design for multi-agent systems encounters challenges associated with unknown uncertainties in agent dynamics. To ensure control performance, linearly parameterized methods have been adopted to estimate gains and approximate unknown nonlinear dynamics in consensus control for various order multi-agent systems[15][16]. However, these methods often rely on the parameterization of uncertainty terms and the a priori knowledge of regressor matrices, satisfying the persistently exciting (PE) condition[17].

As an alternative, neural networks (NN), acting as universal approximators independent of prior knowledge, have gained popularity for compensating unknown uncertainties[18]. While fruitful results have been achieved in consensus or formation control for multi-agent systems. Then, extending these existing results to address the consensus-based formation tracking problem for multi-agent systems proves due to the expanded dimensions and complexities associated with formation control design and analysis[19][20].

To the best of our knowledge, simultaneous recording of state transformation by observers for both the leader and followers often introduces errors, thereby posing challenges to the stability of the system. To address this issue, radial basis function neural networks (NN) are employed to analyze, budget, and supplement the observed errors. The integration of these NNs serves the ultimate goal of ensuring the stability of the system, enabling it to operate reliably.

Building upon the considerations outlined above, this paper introduces an innovative tracking control scheme for multi-agent systems, leveraging neural network (NN) technology. The key contributions of this proposed scheme are summarized as follows:

(i) Utilization of Neural Network Control (NN) to Address Model Uncertainty: The scheme employs Radial Basis Function Neural Network (RBFNN) technology to effectively handle model uncertainty within multi-agent systems. This application of NNs enhances the system's adaptability and robustness.

(ii) Position and Speed Measurement through Observers: A crucial aspect of the proposed scheme involves utilizing observers to measure the position and speed of each agent. This enables the estimation of the states of both the leader and followers. It is noteworthy that achieving system stability is of paramount importance, and the paper emphasizes the need for in-depth analysis and solutions to ensure stable system operation.

(iii) Validation through Simulation Scenarios: To validate the effectiveness of the proposed control scheme, the paper conducts several simulation scenarios. These simulations serve as visual demonstrations, showcasing the performance of the proposed scheme in comparison to previous results and highlighting its superiority. The visual representation of the control scheme in action provides insights into its practical applicability and performance under various conditions.

## 2 Preliminaries and Problem Statements

### 2.1 Preliminaries

*A. Graph Theory*

The connected graph $G = \{V, A, E\}$ is used to describe the communication of a MAS with $N$ agents ($N \in \mathbb{N}^*$), where $V = \{v_1, v_2, \dots, v_N\}$ is the node-set, $A = [a_{ij}] \in \mathbb{R}^{N \times N}$ is the weighted adjacency matrix, and $E = \{e_{ij} = (v_i, v_j): a_{ij} \neq 0, i \neq j, i,j \in \{1,2,\dots,N\}\}$ is the edge set. The in-degree matrix is determined by $D = diag\{\deg_{in}(v_i), i = 1,2,\dots,N\}$, where $\deg_{in}(v_i)$ $(1,2,\dots,N)$ denotes the in-degree of the node $v_i$, which is specified by $\deg_{in}(v_i) = \sum_{j=1}^{N} a_{ij}$. The Laplacian matrix is defined as $L = D - A$.

$$L = diag\{\sum_{j=1}^{N} a_{1j}, \dots, \sum_{j=1}^{N} a_{nj}\} - A \quad (1)$$

It is called to have a directed path from the node $v_i$ to node $v_j$, if there exists a finite number of nodes $(v_{n_1}, v_{n_2}, \dots, v_{n_l})$ $(l \leq N, l \in \mathbb{N}, n_1, n_2, \dots, n_l \in \{1,2,\dots,N\})$ satisfying that $\{e_{i,n_1}, e_{n_1,n_2}, \dots, e_{n_l,j}\} \in E$. A directed graph contains a spanning tree if a node $v_i$ $(i \in \{1,2,\dots,N\})$ has directed paths to any node in the graph.

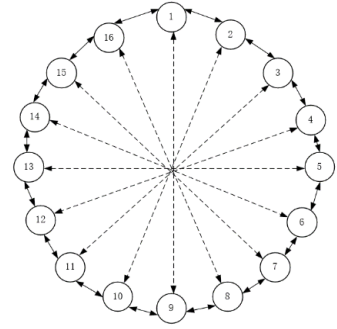

Fig. 1: Directed interaction topology

**Lemma 1.** Let $L \in \mathbb{R}^{N \times N}$ denote the Laplacian matrix of a directed graph $G$. Then, one has:
1) If $G$ is connected, then zero is a simple eigenvalue of $L$, and all the other $N - 1$ eigenvalues are positive $a_{ij} > 0$.
2) The matrix $L$ is symmetric and positive semi-definite, that is, $L^T = L, L \geq 0$.

$$L = \begin{bmatrix} L_{11} & 0 & \dots & 0 \\ L_{12} & L_{22} & \dots & 0 \\ \vdots & \vdots & \ddots & \vdots \\ L_{n1} & L_{n2} & \dots & L_{nn} \end{bmatrix} \quad (2)$$

**Lemma 2.** Let $D \in \mathbb{R}^{N \times N}$ and $A \in \mathbb{R}^{N \times N}$ be the in-degree matrix and adjacency matrix of an undirected graph $G$, respectively. Then, one has:
1) The adjacency matrix A is symmetric;
2) $\forall i \in N^*, (D^{-1}A)^i \geq 0, and\ (D^{-1}A)^{i+1} \geq 0$;
3) The sequence of matrices $(D^{-1}A)^k$ converges to a constant matrix as $k \to \infty$.

## B. Neural Networks

As a general tool to approximate smooth functions, neural networks (NNs) have three parts, including the input layer $(x_i, v_i)$, the hidden layers of $h$ ($h \in N^*$) neurons and the output layer $u_i$. It can be characterized by a weight vector $W$ and other variables such as neural network basic function $\Phi$ and an approximation error $b$.

A smooth function $g(\chi): R^k \to R^l$ ($k, l \in N^*$) can be approximated by a neural network of this type, if the input vector $\chi$ is restricted to a compact set $S \subseteq R^k$. The approximation can be denoted by

$$g(\chi) = W^T \Phi(\chi) + b(\chi) \quad (3)$$

where $W \in \mathbb{R}^h$, $\Phi(\cdot) \in \mathbb{R}^h$ consists of the weight vector of the neural network and its basic function, and the estimated error is limited by a positive constant, that is, $\forall X \in \mathbb{R}^k, |b(\chi)| \leq b_M$.

**Lemma 3.** Let $g(\xi_i, t)$ represent a smooth function. $g(\xi_i, t)$ can be approximated by a neural network with $h$ neurons in its hidden layer.

$$g(\xi_i, t) = W_i^T \Phi(\xi_i, t) + b_i \quad (4)$$

where $\Phi(\xi_i, t) \in \mathbb{R}^h$ is the basic function, $W_i = arg \min_{w \in \mathbb{R}^{q \times h}} \{\sup_{x \in \Omega} \|f(x) - W_i^T \Phi\|\}$, $W_i \in \mathbb{R}^h$ is the weight vector, and $b_i \in \mathbb{R}$ is the error of estimation, which is bounded by a constant $b_M$ ($b_M \in \mathbb{R}, b_M \geq 0$).

## 2.2 Problem Statements

Consider a Second-order nonlinear $N$-agent ($N \in \mathbb{N}^*$) system. The dynamics of the $i^{th}$ agent ($i \in \{1, 2, \cdots, N\}$) can be represented as follows:

$$\begin{cases} \dot{x}_i(t) = v_i(t) \\ \dot{v}_i(t) = f(x_i, v_i) + u_i(t) \end{cases} \quad (5)$$

where $x_i(t) = [x_{i1}, \ldots, x_{iN}]^T \in \mathbb{R}$ is the position, $v_i(t) = [v_{i1}, \ldots, v_{iN}] \in \mathbb{R}$ is the velocity state of agents, and $f(\xi_i, t) \in \mathbb{R}^N$ is a hypothetical smooth and bounded nonlinear function that is unknown, $and\ u_i \in \mathbb{R}^N$ is the control input.

The desired reference signals are described by the following dynamics which is viewed as an independent virtual leader agent,

$$\begin{cases} \dot{\bar{x}}(t) = \bar{v}(t) \\ \dot{\bar{v}}(t) = h(t) \end{cases} \quad (6)$$

where $\bar{x} \in \mathbb{R}^N$ is the reference position trajectory, $\bar{v} \in \mathbb{R}^N$ is reference velocity, $h(\cdot)$ is a smooth bounded function that $h_i(t) = [h_{ix}(t), h_{iv}(t)]^T$.

**Theorem 1**: The second-order leader-follower formation is achieved if the solutions of the multi-agent system satisfy the following conditions.

$$\begin{cases} \lim_{t \to \infty} \|x_i(t) - \bar{x}(t) - p_i\| = 0 \\ \lim_{t \to \infty} \|v_i(t) - \bar{v}(t)\| = 0 \end{cases} \quad (7)$$

where $i = 1, 2, \ldots, N$, $p_i = [p_{i1}, \ldots, p_{iN}]^T \in \mathbb{R}^N$ is a constant vector, which describes the desired relative position between agent $i$ and the leader.

**Control protocol design**: Define the following coordinate transformations

$$\begin{cases} z_{xi}(t) = x_i(t) - \bar{x}(t) - p_i \\ z_{vi}(t) = v_i(t) - \bar{v}(t) \end{cases} i = 1, 2, \ldots, N \quad (8)$$

According to (3) and (4), the following error dynamics are yielded:

$$\begin{cases} \dot{z}_{xi}(t) = z_{vi}(t), \\ \dot{z}_{vi}(t) = u_i + f(x_i, v_i) - h(t) \end{cases} i = 1, 2, \ldots, N \quad (9)$$

Rewrite the error dynamic (7) to the following compact form:

$$\dot{z}(t) = \begin{bmatrix} z_v(t) \\ u + F(z) - h(t) \otimes 1_N \end{bmatrix} \quad (10)$$

where $z(t) = [z_x^T(t), z_v^T(t)]^T \in \mathbb{R}^{N \times D}$ with $z_x(t) = [z_{x1}^T, \ldots, z_{xN}^T]^T \in \mathbb{R}^{N \times D}$ and $z_v(t) = [z_{v1}^T, \ldots, z_{vN}^T]^T \in \mathbb{R}^{N \times D}$, $u = [u_1^T, \ldots, u_N^T]^T \in \mathbb{R}^{N \times D}$, $F(z) = [f_1^T, \ldots, f_n^T]^T \in \mathbb{R}^{N \times D}$. $\otimes$ is a Kronecker product.

Define the position and velocity formation errors as:

$$e_{xi}(t) = \sum_{j \in \Lambda_i} a_{ij}(x_i(t) - p_i - x_j(t) + p_j) \\ + d_i(x_i(t) - \bar{x}(t) - p_i) \quad (11)$$

$$e_{vi}(t) = \sum_{j \in \Lambda_i} a_{ij}(v_i(t) - v_j(t)) + d_i(v_i(t) \\ - \bar{v}(t)), i = 1, \cdots, m \quad (12)$$

Equations (11) and (12) can be rewritten as follows:

$$\begin{cases} e_{xi}(t) = \sum_{j \in \Lambda} a_{ij}(z_{xi}(t) - z_{xj}(t)) + d_i z_{xi}(t), \\ e_{vi}(t) = \sum_{j \in \Lambda} a_{ij}(z_{vi}(t) - z_{vj}(t)) + d_i z_{vi}(t) \end{cases} \quad (13)$$

$$i = 1, 2, \ldots, N$$

where $e(t) = [e_x^T(t), e_v^T(t)]^T \in \mathbb{R}^{N \times D}$ with $z_x(t) = [z_{x1}^T, \ldots, z_{xN}^T]^T \in \mathbb{R}^{N \times D}$ and $z_v(t) = [z_{v1}^T, \ldots, z_{vN}^T]^T \in \mathbb{R}^{N \times D}$, $d = [d_1, \ldots, d_N]^T \in \mathbb{R}^{N \times D}$.

## 3 Main Results

In this section, a nonlinear protocol is taken into consideration, and competent NN structures are suggested for MAS to be effective in achieving group formation. The nonlinear term in the control protocol can be approximated by neural networks as shown in Fig.1.

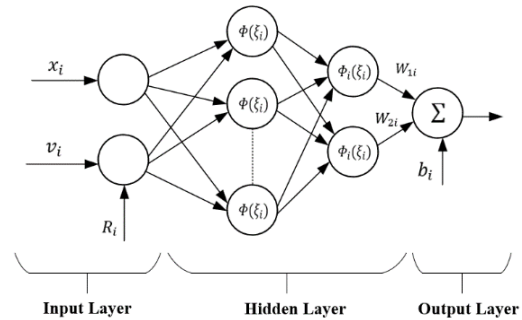

Fig.2: Structure of recurrent neural networks

In (11), the nonlinear function $f_i(x_i, v_i)$ is unknown but continuous, give a compact set $\Omega_i \in \mathbb{R}^2$, for $[x_i^T, v_i^T]^T \in$

$\Omega_i$. it can be re-described by using its NN approximation as:

$$f_i(x_i, v_i) = W_i^{*T}\Phi_i(x_i, v_i) + b_i(x_i, v_i) \qquad (14)$$

Where $W_i^* \in \mathbb{R}^{q_i \times N}$ is the ideal NN weight matrix with the NN neuron number $q_i$, $\Phi_i(x_i, v_i) \in \mathbb{R}^{q_i}$ is the basis function vector, $b_i(x_i, v_i) \in \mathbb{R}^N$ is the approximation error satisfied $\|b_i(x_i, v_i)\| \leq \delta_i$, where $\delta_i$ is a constant.

In (9), consider the control protocol $u_i$ for each agent $i$ ($i \in \{1, 2, \cdots, N\}$) In (2), since the ideal weight matrix $W_i^*$ is an unknown constant matrix, it is unavailable for the actual control design. By using the estimation $\widehat{W}_i(t)$ of the ideal NN weight $W_i^*$, the formation control is constructed in the following:

$$u_i(t) = -\gamma_x e_{xi}(t) - \gamma_v e_{vi}(t) - \widehat{W}_i^T(t) \times \Phi(x_i, v_i) \qquad (15)$$

$$i = 1, 2, \ldots, N$$

where $\gamma_x > 0, \gamma_v > 0$ are two design constants $\widehat{W}_i(t) \in \mathbb{R}^{m \times N}$ is the estimation of $W_i^*$.

The NN updating law for tuning $W_i^*$ is given in the following:

$$\dot{\widehat{W}}_i(t) = \Gamma_i \left( \Phi_i(x_i, v_i)(e_{xi}(t) + e_{vi}(t))^T - \sigma_i \widehat{W}_i(t) \right) \qquad (16)$$

$$i = 1, 2, \ldots, N$$

where $\Gamma_i \in \mathbb{R}^{m \times m}, m = N - 1$ is a positive definite matrix, $\sigma_i > 0$ is a design constant.

**Theorem 2**. Lyapunov Stability Theory
Consider a dynamical system which satisfies

$$\begin{cases} \dot{x}_i(t) = v_i(t) \\ \dot{v}_i(t) = f(x_i, v_i) + u_i(t) \end{cases}, t > 0 \qquad (17)$$

The equilibrium point $x^* = 0$ is stable and $f(x_0, v_0) = \varepsilon_i(0)$, if there is a scalar function $V(x)$ which has a continuous first-order derivative, and satisfies $V(0) = 0$ any non-zero point $x$ in the state space satisfies:

1. $V(x)$ is positive;
2. $\dot{V}(x)$ is negative;
3. When $t \to \infty, V \to \infty$;

So, if the origin of the system is balanced in a large range of increasingly stable, the system is stable.

For the non-linear second-order multi-agent system, if the system formation control with NN weight updating law is performed for the system and can design constants are chosen to satisfy

$$\gamma_x > 1, \gamma_v > 1.5 + 1/2(\lambda_{\min}^L)^2$$

$$\gamma_x + \gamma_v > \frac{1}{\lambda_{\min}^L} \qquad (18)$$

where $\lambda_{\min}^L$ is the minimal eigenvalue of matrix $L$, then the following control objectives can be achieved.

The Lyapunov theory will be used to prove that the non-linear multi-agent system is stable:

$$V(t) = \frac{1}{2} z^T(t) \left( \begin{bmatrix} (\gamma_x + \gamma_v)\hat{L}\hat{L} & \hat{L} \\ \hat{L} & \hat{L} \end{bmatrix} \otimes I_N \right) \\ \times z(t) + \frac{1}{2} \sum_{i=1}^N \text{Tr}\{\widetilde{W}_i^T(t)\Gamma_i^{-1}\widetilde{W}_i(t)\}, \qquad (19)$$

where $\hat{L} = L + D$, and $\widetilde{W}_i(t) = \widehat{W}_i(t) - W_i^*$.

Given equation (15) the fact $(\gamma_x + \gamma_v)\hat{L}\hat{L} - \hat{L} > 0$, hence $\begin{bmatrix} (\gamma_x + \gamma_v)\hat{L}\hat{L} & \hat{L} \\ \hat{L} & \hat{L} \end{bmatrix}$ is a positive definite matrix, thus the function $V(t)$ can be considered as a Lyapunov function candidate, and $V(t) > 0$.

The time derivative of $V(t)$ is:

$$\dot{V}(t) = z^T(t) \left( \begin{bmatrix} (\gamma_x + \gamma_v)\hat{L}\hat{L} & \hat{L} \\ \hat{L} & \hat{L} \end{bmatrix} \otimes I_N \right) \\ \times \begin{bmatrix} z_v(t) \\ u + F(z) - h(t) \otimes 1_N \end{bmatrix} \\ + \sum_{i=1}^N \text{Tr}\{\widetilde{W}_i^T(t)(\Phi_i(x_i, v_i) \\ \times (e_{xi}(t) + e_{vi}(t))^T - \sigma_i \widetilde{W}_i(t))\} \qquad (20)$$

Using the facts $e_x(t) = \hat{L} z_x(t)$ and $e_v(t) = \hat{L} z_v(t)$, then the equation can be written as

$$\dot{V}(t) = \sum_{i=1}^N \left( (\gamma_x + \gamma_v) e_{xi}^T(t) e_{vi}(t) + e_{vi}^T(t) \right. \\ \times z_{vi}(t) + \sum_{i=1}^N (e_{xi}^T(t) + e_{vi}^T(t)) \\ \times (u_i + f_i(x_i, v_i) - h(t)) \\ + \sum_{i=1}^N \text{Tr}\{\widetilde{W}_i^T(t)(\Phi_i(x_i, v_i)(e_{xi}(t) \\ + e_{vi}(t))^T - \sigma_i \widetilde{W}_i(t))\} \qquad (21)$$

Inserting the NN approximation and the controller into (16) yields

$$\dot{V}(t) = \sum_{i=1}^N \left( (\gamma_x + \gamma_v) e_{xi}^T(t) e_{vi}(t) \right. \\ + e_{vi}^T(t) z_{vi}(t)) + \sum_{i=1}^N (e_{xi}^T(t) \\ + e_{vi}^T(t))(-\gamma_x e_{xi}(t) - \gamma_v e_{vi}(t) \\ - \widehat{W}_i^T(t)\Phi_i(x_i, v_i) + W_i^{*T}\Phi_i(x_i, v_i) \\ + b_i(x_i, v_i) - h(t)) \\ + \sum_{i=1}^N \text{Tr}\{\widetilde{W}_i^T(t)\Phi_i(x_i, v_i)(e_{xi}(t) \\ + e_{vi}(t))^T - \sigma_i \widetilde{W}_i^T(t)\widetilde{W}_i(t)\}. \qquad (22)$$

The equation $\widehat{W}_i(t) = \widetilde{W}_i(t) - W_i^*$, we get

$$\dot{V}(t) = -\sum_{i=1}^N \gamma_x e_{xi}^T(t) e_{xi}(t) - \sum_{i=1}^N \gamma_v \\ \times e_{yi}^T(t) e_{vi}(t) + \sum_{i=1}^N e_{vi}^T(t) z_{vi}(t) \\ - \sum_{i=1}^N (e_{xi}^T(t) + e_{vi}^T(t))\widehat{W}_i^T(t) \\ \times \Phi_i(x_i, v_i) + \sum_{i=1}^N (e_{xi}^T(t) + e_{vi}^T(t)) \\ \times b_i(x_i, v_i) - \sum_{i=1}^m (e_{xi}^T(t) + e_{vi}^T(t)) \\ \times h(t) + \sum_{i=1}^N Tr\{\widetilde{W}_i^T(t)\Phi_i(x_i, v_i) \\ \times (e_{xi}^T(t) + e_{vi}^T(t))\} - \sum_{i=1}^m \text{Tr}\{\sigma_i \\ \times \widehat{W}_i^T(t)\widetilde{W}_i(t)\} \qquad (23)$$

Then, the following inequality can be derived:

$$\dot{V}(t) \leq -\left(\frac{\lambda_{\min}^a}{\lambda_{\max}^b}\right) z^T(t) \left(\begin{bmatrix} (\gamma_x + \gamma_v)\hat{L}\hat{L} & \hat{L} \\ \hat{L} & \hat{L} \end{bmatrix} \otimes I_N\right) z(t) - \frac{1}{2}\sum_{i=1}^{m} \frac{\sigma_i}{\Gamma_i^{-1}} T_r\left\{\widehat{W}_i^T(t) \times \Gamma_i^{-1}\widehat{W}_i(t)\right\} + c \quad (24)$$

where $\lambda_{\min}^a$ is the minimal eigenvalue of matrix $a$, and $\lambda_{\min}^b$ is the maximal eigenvalue of matrix $b$.

The above inequality can be rewritten as:

$$\dot{V}(t) \leq -aV(t) + c$$

where $a$ and $c$ are constants, when $t \to \infty, \dot{V}(t) \to \infty$, then $V(t) \to \infty$. From (15) and (17), it can be concluded that the second-order multi-agent system with the formation protocol (12) and NN weight law (13) updating is stable.

## 4 Numerical Simulations

Consider a MAS containing 16 agents that are characterized by:

$$\begin{cases} \dot{x}_i(t) = v_i(t) \\ v_i(t) = u_i + \begin{bmatrix} x_{i1} + \alpha_i \cos^2(x_{i1}v_{i1}) \\ v_{i2} + \beta_i \sin^2(x_{i2}v_{i2}) \end{bmatrix} \end{cases}$$

$$i = 1,2,\ldots,16$$

where $x_i(t) = [x_{i1}, x_{i2}]^T$ is the position in $(x,y)$ direction, $v_i(t) = [v_{i1}, v_{i2}]^T$ is the velocity in $(x,y)$ direction, $\boldsymbol{\alpha} = [\alpha_1, \alpha_2, \ldots, \alpha_{16}]^T$ and $\boldsymbol{\beta} = [\beta_1, \beta_2, \ldots, \beta_{16}]^T$.

The initial values are
$x_i(0) = [4,6]^T, [5.5,1]^T, [2.5,5]^T, [8,2]^T, [6.5,5.5]^T$
$[1.5,4]^T, [1.5,6.5]^T, [5,6]^T, [2,7.5]^T, [6,5]^T, [9,7.5]^T$
$[4,5]^T, [3.5,4]^T, [5,3.5]^T, [1.5,2.5]^T, [4,7]^T$

$v_i(0) = [5,5]^T, [2,3.5]^T, [1,5.5]^T, [2,6.5]^T, [7,2]^T$
$[2.5,3.5]^T, [3,5]^T, [4,5.3]^T, [5,5.7]^T, [1,2.5]^T, [4,5.5]^T$
$[4.5,5]^T, [6,4.5]^T, [4,5.6]^T, [7,3.5]^T, [2,4]^T$

$\boldsymbol{\alpha} = [0.5, 0.3, -0.2, 0.7, -0.5, -0.3, -0.6, 0.4, -0.8, 0.6,$
$-0.4, 0.3, 0.4, -0.2, 0.7, -0.6]^T$

$\boldsymbol{\beta} = [0.7, 0.35, -0.2, 0.6, -0.25, -0.5, 0.45, -0.4, 0.75, -0.6,$
$-0.3, -0.4, 0.65, -0.25, 0.8, -0.45]^T]$

**Summary Simulation:** To demonstrate the theoretical results derived in the preceding sections, numerical simulations were conducted in two distinct scenarios.

In Scenario 1, a simulation was crafted with specific data values for the system, encompassing parameters such as initial position and velocity. A neural network was employed to compute the data and regulate the system based on the provided parameters.

In Scenario 2, a simulation with practical applications was executed, taking into account disturbances affecting the system. Here, a multi-agent system was represented by sixteen unmanned vehicles, illustrating the formation they collectively create. This application-based simulation offered insights into the system's behavior under real-world conditions, showcasing the effectiveness of the proposed control scheme in forming and maintaining a desired configuration despite external disturbances.

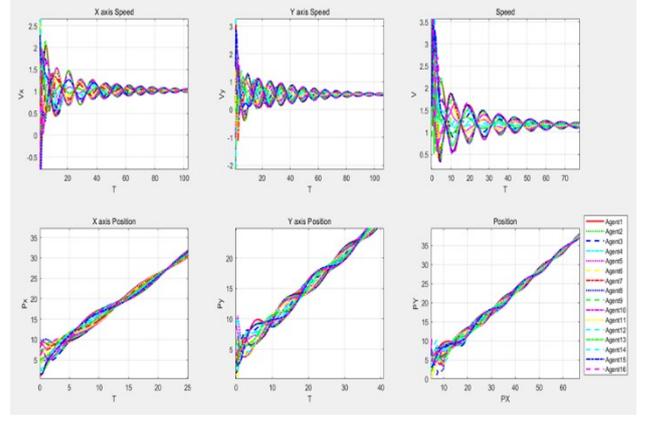

Fig.3: Convergence curves of the velocity and location of the 16-agent MAS

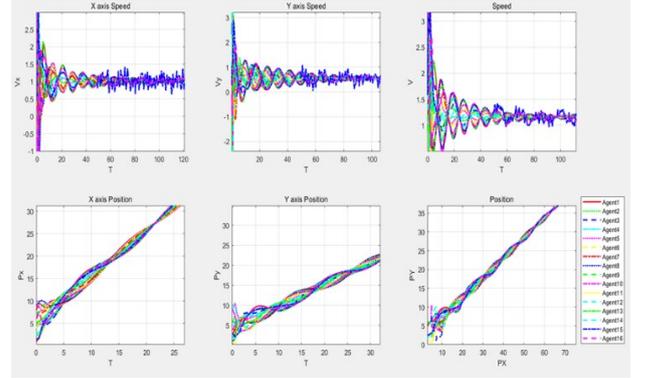

Fig.4: Convergence curves of the velocity and location of the 16-agent MAS (interference condition)

For fig-3, Simulation results demonstrate the effectiveness of the proposed multi-agent system in measuring the position and velocity of each agent. The analysis of position and speed transformation states emphasizes the gradual transformation between different positions and speeds to achieve consistent system operation. In addition, the simulation results also show that the system can operate stably.

For fig-4, Simulation results represents When the system is interference, some agents or the entire system is disturbed, causing errors in their speeds. However, after correction through the system's neural network, the system's position is not affected and it can operate and form consistently. And fig-5 Sixteen agents successfully formed and maintained a regular hexagonal formation with identical edges while also operating consistently.

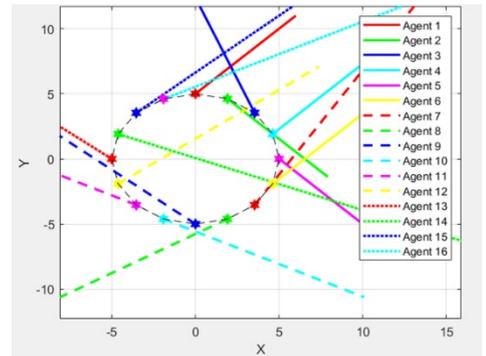

Fig.5: A local view of the MAS formation

## 5 Conclusion

This paper successfully accomplished the cooperative formation of multiple agents characterized by nonlinear dynamics through the utilization of neural networks. The derivation of sufficient conditions for achieving Multi-Agent System (MAS) formation control constitutes a significant contribution. Numerical simulation examples were presented to validate the obtained results. The simulations clearly illustrated that the proposed distributed control protocol guarantees consistency and accuracy in the cooperative formation of MAS.

Furthermore, there is potential for enhancing the proposed formation control algorithm by incorporating more sophisticated neural network structures and optimization methods. This avenue for improvement could lead to enhanced performance and adaptability, offering opportunities for refining the system's capabilities in handling complex scenarios and dynamic environments. The results obtained in this study provide a foundation for future advancements and optimizations in the realm of cooperative formation control for multi-agent systems.